\documentclass[pmlr]{jmlr}

\usepackage{booktabs}
\usepackage{placeins}
\usepackage{tikz}
\usepackage{pgfplots}
\pgfplotsset{compat=1.18}
\usetikzlibrary{positioning,arrows.meta,fit,calc,shapes.misc}

\tikzset{
  pbox/.style={draw=black!60, fill=black!4, rounded corners=2pt, align=center,
               font=\small, inner sep=5pt, minimum height=9mm},
  ptitle/.style={font=\small\itshape, text=black!65},
  parrow/.style={-{Latex[length=2mm]}, draw=black!70, thick},
  plabel/.style={font=\footnotesize, text=black!65, fill=white, inner sep=1.5pt},
}
\definecolor{caught}{gray}{0.25}
\definecolor{missed}{gray}{0.55}
\definecolor{unk}{gray}{0.82}
\definecolor{blameful}{gray}{0.25}
\definecolor{blameless}{gray}{0.68}
\pgfplotsset{
  stratbars/.style={
    ybar, bar width=9pt, height=4.6cm, ymin=0, clip=false,
    enlarge x limits=0.3,
    symbolic x coords={Sonnet 4.6, Opus 4.7, GPT-5.5}, xtick=data,
    x tick label style={font=\footnotesize, rotate=20, anchor=north east},
    y tick label style={font=\footnotesize},
    title style={font=\small, yshift=1mm},
    nodes near coords, nodes near coords style={font=\scriptsize},
    axis lines*=left, ymajorgrids, grid style={black!12},
    every axis plot/.append style={draw=none},
    legend style={font=\small, draw=none, legend columns=2,
      /tikz/every even column/.append style={column sep=6pt}},
    legend image code/.code={\fill[#1] (0cm,-0.1cm) rectangle (0.3cm,0.15cm);},
  },
}
\pgfplotsset{
  verdictbars/.style={
    ybar, bar width=11pt, width=\linewidth, height=5.4cm, clip=false,
    ymin=0, enlarge x limits=0.25,
    symbolic x coords={GPT-5.5, Sonnet 4.6, Opus 4.7}, xtick=data,
    ylabel={Instances}, ylabel style={font=\small},
    tick label style={font=\small}, title style={font=\small, yshift=5mm},
    legend style={font=\small, at={(0.5,-0.24)}, anchor=north, legend columns=3, draw=none, /tikz/every even column/.append style={column sep=6pt}},
    legend image code/.code={\fill[#1] (0cm,-0.1cm) rectangle (0.3cm,0.15cm);},
    nodes near coords, nodes near coords style={font=\footnotesize},
    axis lines*=left, ymajorgrids, grid style={black!12},
    every axis plot/.append style={draw=none},
  },
}
\newcommand{\verdictchart}[8]{%
\begin{tikzpicture}
\begin{axis}[verdictbars, title={#1}, ymax=#2, ytick distance=10]
\addplot[fill=caught] coordinates {#3};
\addplot[fill=missed] coordinates {#4};
\addplot[fill=unk]    coordinates {#5};
\legend{Caught (yes), Missed (no), Unknown}
\node[font=\small\bfseries, anchor=south] at (axis cs:GPT-5.5, #2)   {Discovery rate #6};
\node[font=\small\bfseries, anchor=south] at (axis cs:Sonnet 4.6, #2) {Discovery rate #7};
\node[font=\small\bfseries, anchor=south] at (axis cs:Opus 4.7, #2)  {Discovery rate #8};
\end{axis}
\end{tikzpicture}}

\jmlrvolume{349}
\jmlryear{2026}
\jmlrworkshop{Conference on Applied Machine Learning for Information Security}

\title[Evaluating LLM Security Review Agents]{Evaluating LLM Security Review
  Agents with Operationally Sourced Ground Truth}

\author{\Name{Mudita Khurana} \Email{mudita.khurana@airbnb.com}\and
  \Name{Karim Liman-Tinguiri} \Email{karim.limantinguiri@airbnb.com}\\
  \addr Airbnb, Inc.}

\editor{Edward Raff and Ethan M. Rudd}

\begin{document}

\maketitle

\begin{abstract}
The growing volume of AI-assisted code has placed new strain on security
review, and although LLM-based review agents can increasingly reason about
vulnerabilities that deterministic tools miss, deploying them responsibly
first requires knowing whether they actually work. Existing benchmarks answer
this question only partially, because they construct ground truth
synthetically or from public CVE databases and are therefore blind to the
organisation-specific vulnerability patterns that dominate a given production
codebase. We present an evaluation framework that addresses these limitations
by grounding evaluation in an organisation's own vulnerabilities, and we make
three contributions: a \textbf{dataset construction methodology} rooted in
operational security data, \textbf{two complementary evaluation strategies},
and a \textbf{trace-level behavioural characterisation} of how frontier
models reason about the vulnerabilities in our corpus. Our \textbf{dataset construction methodology} derives
ground truth from operational security data, exploiting the temporal property
that a confirmed vulnerability must exist immediately before its fix is merged
and be absent immediately after, and it draws on bug bounty reports,
penetration test findings, and internally reported vulnerabilities. The \textbf{two complementary evaluation strategies} pose different questions: one reviewing
the pull request that introduced a vulnerability and one searching a
repository where it already exists. Both strategies run through a two-agent
pipeline, pairing a Security Engineer agent that reviews code without
knowledge of the underlying vulnerability with a Judge agent that decides
whether its findings would have mitigated the known issue. Across three
frontier models (GPT-5.5, Sonnet~4.6 and Opus~4.7), GPT-5.5 shows higher
discovery rates on the pull-request strategy (57\%, 95\%~CI [42\%, 72\%])
after a round of OPRO-style improvements, and Sonnet~4.6 on the repository
strategy (69\%, 95\%~CI [56\%, 81\%]). Going beyond these outcome scores, our \textbf{trace-level behavioural characterisation} shows that the
models exhibit
complementary strengths that explain these differences, and that outcome-only
scoring can mischaracterise their production behaviour. The methodology relies only on artefacts that vulnerability management
workflows commonly produce, which makes it adaptable by other organisations
with a similar workflow, though all empirical results in this paper are drawn
from a single organisation's codebase and we have not yet tested how well they
generalise.
\end{abstract}


\FloatBarrier
\section{Introduction}
\label{sec:intro}

The pace of software development has accelerated significantly with the
adoption of AI-assisted coding tools, a trend widely observed across the
industry in recent years \citep{daniotti2025}. Empirical studies confirm that
LLM-generated code contains security weaknesses at measurable rates across
real production repositories \citep{fu2023}, and the volume of AI-assisted
commits means the traditional model of manual security review, even when
augmented by deterministic static analysis, is under increasing strain
\citep{perry2023}. Deterministic tools remain a critical gate in the CI/CD
pipeline, but they are fundamentally pattern-based and find only what they
have been programmed to find. They miss vulnerability classes that require
contextual reasoning, understanding of authorisation semantics, or knowledge
of how components interact across service boundaries \citep{yu2024}.
LLM-based security agents address this gap directly: an agent that can browse
a repository, trace data flows across files, and reason about whether a code
change introduces an exploitable vulnerability brings a capability that no
static analysis rule can replicate \citep{bigsleep2024}. Deploying such agents
responsibly, however, first requires knowing whether they actually work.

Establishing this is harder than it appears: whether an organisation is
evaluating a commercial security review tool or building one internally, the
natural first step is to test against an existing benchmark. The problem is
that today's most widely used benchmarks, including
CyberSecEval \citep{bhatt2023}, SecRepoBench \citep{shen2025}, and
CWEval \citep{peng2025}, construct ground truth synthetically or from public
CVE databases. They measure performance on generic vulnerability classes (SQL
injection, unsafe deserialisation, standard authentication bypasses, etc.) and
are blind to the patterns that dominate any specific organisation's production
codebase: authorisation gaps in proprietary frameworks, misuse of internal
libraries, or data exposure bugs that only surface in the context of a
specific service architecture. A system that performs well on generic
benchmarks may consistently miss the most prevalent vulnerability class in
your production environment \citep{shen2025}. The most direct way to know whether
your security review agent will work on your code is to evaluate it against
your vulnerabilities, which means constructing your own ground truth dataset.
Yet building one requires solving a problem the community has barely
addressed: how do you establish, programmatically and at scale, that a
specific vulnerability existed at a specific point in a codebase?
Organisations that have tried have done so in isolation, unable to share their
vulnerability data, and when those teams move on, the methodology disappears
with them. No shared practitioner foundation exists for this problem.

Even when evaluation datasets do exist, there is a third challenge that goes
largely unacknowledged. If agent~A catches 52\% of your vulnerabilities and
agent~B catches 32\%, you know which to deploy today, but you do not know why
the gap exists, whether it is closeable, or how to get from 52\% to 70\%
tomorrow without resorting to costly trial and error \citep{khurana2025}.
Outcome scores do not tell you whether the agent is failing because it does
not look in the right places, because it looks but reasons incorrectly, or
because it produces valid observations framed in ways that do not translate to
actionable security output. These are different problems requiring different
interventions, and a single discovery rate number cannot distinguish between
them. Without behavioural insight into how a system fails, improving it is
guesswork.

This paper addresses all three challenges. (i)~We construct evaluation ground
truth directly from operational security data by exploiting a temporal
property that every confirmed, code-rooted vulnerability carries: it must
exist in the codebase immediately before its fix merges and be absent
immediately after. This double-sided oracle lets us derive a labelled dataset
from the same bug bounty reports, penetration test findings, and internally
reported vulnerabilities that many security teams already track, without
synthetic construction or public CVE data. (ii)~We describe the methodology in enough operational detail that
organisations with a similar vulnerability management workflow can adapt it to
their own data, although we have validated it in only one organisation. (iii)~We apply two complementary evaluation strategies across three
frontier models, going beyond outcome scores by conducting a trace-level
behavioural analysis that reveals how these systems actually fail and what
that means for deployment decisions. The remainder of the paper is organised
as follows. Section~\ref{sec:related} discusses related work.
Section~\ref{sec:contrib} summarises our contributions.
Section~\ref{sec:dataset} describes dataset construction.
Section~\ref{sec:pipeline} describes the evaluation pipeline.
Section~\ref{sec:experiments} presents experiments, results, and behavioural
analysis. Section~\ref{sec:future} discusses future work and
Section~\ref{sec:conclusion} concludes.

\section{Related Work}
\label{sec:related}

\paragraph{Security of LLM-generated code.}
\citet{fu2023} conducted a large-scale study of weaknesses in code generated
by Copilot and similar tools across real GitHub projects, finding that 29.5\%
of Python snippets examined contained security weaknesses spanning 43~distinct
CWE categories. Earlier controlled studies reached similar conclusions, with
\citet{pearce2022} reporting that roughly 40\% of Copilot completions were
vulnerable, and \citet{perry2023} showing that developers using AI assistants
not only introduced more security bugs but were also more likely to believe
their code was secure. These findings ground the central concern motivating
our work: if AI-assisted generation introduces vulnerabilities at measurable
rates and the volume of such code continues to climb, the security review
systems responsible for catching those vulnerabilities face a scale and
coverage challenge that rule-based approaches alone cannot meet. Our work does
not assess code generation models, but rather the downstream agents
responsible for reviewing what those models produce.

\paragraph{LLM-based security review and benchmarks for secure code.}
Beyond generation, LLMs have been studied directly as security reviewers, and
\citet{yu2024} found that across seven models they consistently outperformed
state-of-the-art static analysis tools at detecting security defects during
code review. To measure capabilities of this kind, several benchmarks have
been proposed. \citet{bhatt2023} established a methodology for
security-focused LLM evaluation, measuring whether capable models emit
insecure code when prompted and providing an early benchmark others have built
upon. \citet{peng2025} showed that earlier benchmarks conflated functional
correctness with security correctness, motivating evaluation frameworks that
treat security as a distinct dimension. \citet{shen2025} demonstrated that
performance on self-contained benchmarks does not transfer to the repository
level, which motivates evaluation approaches that derive ground truth from an
organisation's own production codebase rather than from generic corpora.
\citet{jimenez2024} established that LLM agents evaluated on GitHub issue
resolution at the repository level perform very differently from agents
evaluated on isolated code snippets, further pointing toward the need for
realistic, repository-grounded evaluation contexts. Closest to our approach,
\citet{chen2026securevibebench} introduce SecureVibeBench, which constructs
realistic secure-coding tasks from open-source vulnerability histories and
evaluates whether code agents produce functionally correct and secure code. Two properties separate this prior work from ours: existing benchmarks
derive ground truth from synthetic tasks, public vulnerability corpora, or
open-source vulnerability histories, and they evaluate code
\emph{generation} rather than vulnerability \emph{detection}.

\paragraph{Constructing ground truth from real vulnerabilities.}
The problem of identifying which change introduced a known defect has a long
history in the defect prediction literature, beginning with the SZZ
algorithm \citep{sliwerski2005}. Our blameful strategy draws on the same
intuition but differs in two respects: we apply it to LLM agent evaluation
rather than defect localisation, and we hold the result to a human security
engineer's manual confirmation. Public datasets such as
CVEfixes \citep{bhandari2021} mine CVE records to assemble before-and-after
versions of vulnerable code across open source projects, while subsequent
audits have shown that labels derived directly from fix commits are frequently
noisy \citep{chen2023noise}. The closest industry parallel is Meta's ACH
system \citep{foster2025}, which deliberately injects faults into internal
production code and generates tests guaranteed to catch them, though ACH
targets fault-specific regression test generation rather than the evaluation
of vulnerability detection agents. Our blameless strategy, which constructs
evaluation instances directly from fix pull requests without any blame
attribution, has no direct analogue in this literature that we are aware of.

\paragraph{Agentic code analysis.}
Our Security Engineer follows the ReAct framework \citep{yao2023}, which
interleaves reasoning traces with tool use in an iterative loop. Agentic
approaches of this kind have begun to demonstrate concrete security value,
with \citet{bigsleep2024} finding a previously unknown exploitable
vulnerability in widely used production software, and related work showing
that coordinated teams of agents can surface and exploit real
vulnerabilities \citep{zhu2024}.

\paragraph{LLM-as-judge evaluation.}
Using LLMs as automated evaluators has become a standard pattern for tasks
where ground truth is difficult to specify formally \citep{zheng2023}. Our
Judge applies this pattern to a security-specific verdict. A growing body of
work has documented that LLM judges carry systematic biases and reliability
limits \citep{gu2024}, and our own experience with judge calibration
reinforces the view that the judge is not a neutral oracle but a learned
system whose accuracy bounds the meaningfulness of any metric it produces.

\paragraph{Prompt optimisation for agents.}
\citet{yang2024} showed that a meta-optimiser model can iteratively improve a
target prompt by examining failure cases without access to model weights. We
adopt the same technique for the Security Engineer prompt. The original OPRO
work already observed that optimised prompts can overfit when no held-out
split is maintained \citep{yang2024}, a concern we see in our own setting.

\FloatBarrier
\section{Contributions}
\label{sec:contrib}

This paper makes three contributions to the evaluation of LLM-based security
code review systems.

\textbf{A ground-truth dataset construction methodology grounded in
operational security data.} We introduce a dataset construction approach that
derives evaluation ground truth directly from confirmed vulnerabilities in
production systems. The methodology exploits a natural temporal property of
every code-rooted vulnerability: it must exist immediately before its fix
merges and be absent immediately after. We use this as a double-sided oracle
to validate dataset quality automatically, providing a principled and
reproducible alternative to synthetic benchmark construction.

\textbf{Two complementary evaluation paradigms for measuring agent
performance.} The blameful strategy presents the agent with the pull request
that introduced a known vulnerability. The blameless strategy presents the
agent with the repository state in which a known vulnerability exists.
Together these two paradigms provide complementary evaluation signals for two
central tasks that a security code review agent performs in production
systems, and we characterise the tradeoffs between them in terms of ecological
validity, dataset scale, and construction cost.

\textbf{A behavioural characterisation of frontier model security reasoning
across evaluation strategies.} We apply our evaluation pipeline to compare
models from two frontier providers and conduct a trace-level behavioural
analysis that goes beyond outcome scores. We find that the two model families
exhibit complementary strengths: one demonstrates stronger attacker-capability
reasoning, completing the chain from an observed code anomaly to a concrete
exploit scenario, while the other demonstrates stronger exploratory search
behaviour, reading more files and using targeted vulnerability-pattern
searches to surface issues the first model drifts past. We observe that the relative performance of models differs across the two
evaluation strategies we used, and that outcome-only scoring can
mischaracterise model capability by failing to account for these behavioural
differences.

\FloatBarrier
\section{Dataset Construction}
\label{sec:dataset}

Building a rigorous ground-truth dataset for evaluating LLM-based security
code review requires solving a problem that existing public benchmarks largely
sidestep: how do you obtain confirmed vulnerabilities that reflect the actual
distribution of bugs your system will encounter in production?
\citet{bhatt2023} and \citet{peng2025} construct ground truth synthetically by
generating vulnerable code to order, while \citet{shen2025} draws from public
CVEs in open-source C/C++ repositories. These approaches are valuable for
measuring generic capability across common weakness classes, but they do not
capture the patterns that dominate a specific organisation's codebase, such as
authorisation gaps in proprietary access control frameworks, authentication
misuse tied to internal libraries, or data exposure bugs that only surface in
the context of a specific service architecture \citep{zhou2025lessleak}.

We took a different approach, deriving ground truth directly from our
operational security data: confirmed vulnerabilities that were found, triaged,
and fixed in production. This gives us ground truth that is, by construction,
representative of the bugs our security code review agent will actually
encounter. The central insight driving our methodology is that every confirmed,
code-rooted vulnerability carries a natural temporal oracle: the vulnerability
must exist in the codebase immediately before its fix is merged, and must be
absent immediately after.

The idea of pairing vulnerable and fixed code drawn from fix commits is not
new in itself. Public datasets such as CVEfixes \citep{bhandari2021} mine CVE
records from the National Vulnerability Database to assemble before-and-after
versions of vulnerable code across open source projects. What distinguishes
our approach is that we treat the before-to-after transition not merely as a
labelling device but as an automated quality oracle, and we apply it to
confirmed vulnerabilities from our own operational security data rather than
to publicly disclosed CVEs. We exploit this oracle in two complementary ways,
which we call the \emph{blameful} and \emph{blameless} evaluation strategies,
described in detail in Sections~\ref{sec:blameful}
and~\ref{sec:blameless} respectively.

\subsection{Data Sources and Filtering}
\label{sec:sources}

Our dataset draws from three sources of confirmed security vulnerabilities,
all tracked as tickets in our internal system.

The first source is our public bug bounty program, where submissions
go through triage, validation, and remediation, with the ticket created at
validation and linked to the fix~PR at remediation, giving us externally
discovered and confirmed vulnerabilities with direct references to the code
that addressed them.

The second source is third-party penetration testing. We conduct
periodic engagements with external security firms whose findings are filed as
tickets following the same structure as bug bounty reports, including fix~PR
references for code-rooted issues.

The third source is internal engineer-reported vulnerabilities.
Security engineers file tickets against production code drawing on static
analysis tooling, manual code review, and internal red team engagements.
These findings are validated by the reporting engineer and tracked through the
same remediation workflow as externally reported issues.

We applied several filters to produce a clean dataset for evaluation. We
excluded all tickets without an associated fix~PR, since the temporal oracle
requires a concrete code change representing the remediation. We also excluded
tickets where the fix does not manifest as an application code change. After
filtering, we observed that authorisation issues were substantially
over-represented relative to other vulnerability classes, a reflection of the
discovery dynamics of our three source channels. We stratified the corpus into
authorisation and non-authorisation subsets and sampled proportionally across
both, a choice that reflects a degree of subjective judgement about what
distribution is most useful for evaluating an agent that will review
production pull requests. We note that other organisations applying this
methodology may reasonably make different sampling choices depending on their
own vulnerability distribution and deployment context.

The filtered corpus forms the shared foundation from which both the blameful
and blameless datasets are derived. The blameful dataset requires manual
validation of each culprit~PR, which is the binding constraint on its size and
results in 40~confirmed instances. The blameless dataset does not require
blame attribution at all, allowing it to include all 49~instances that pass
the temporal oracle quality filter. Figure~\ref{fig:pipeline-dataset}
illustrates this construction process. This size asymmetry is a direct
consequence of the tradeoff between ecological validity and scalability
described in Section~\ref{sec:comparison}.

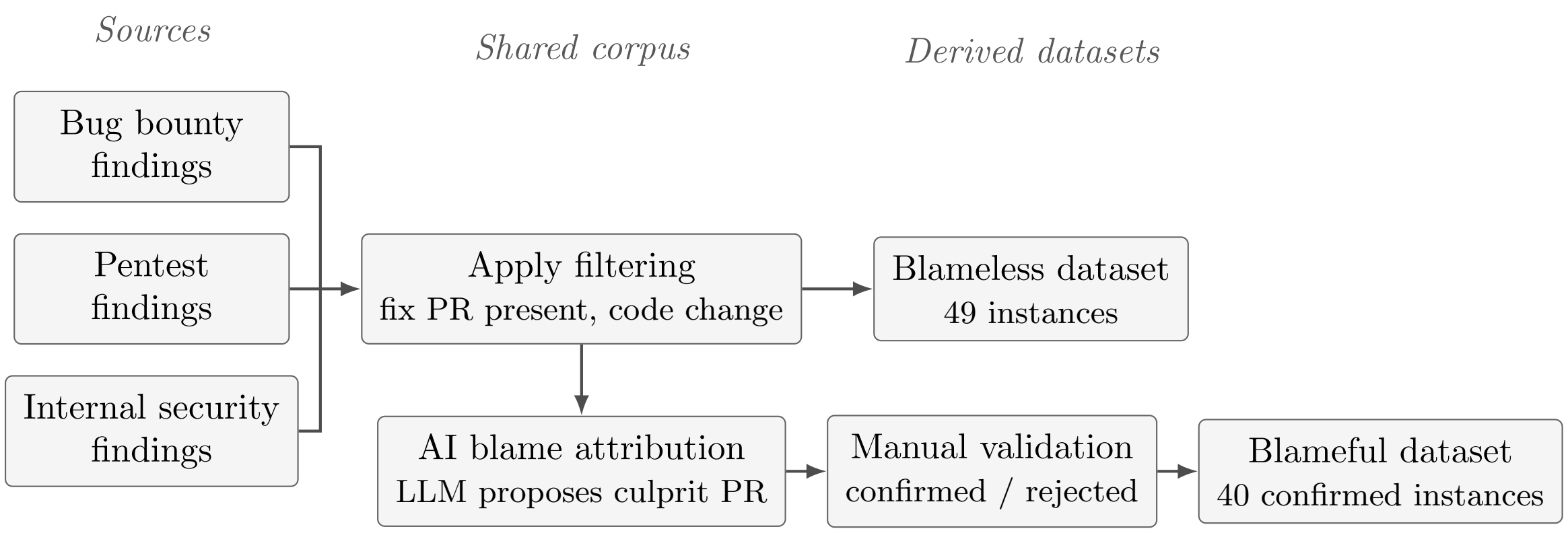
\begin{figure}[htbp]
\floatconts
  {fig:pipeline-dataset}%
  {\caption{Dataset construction pipeline. The three source channels converge
    through a shared filtering step, after which the corpus diverges into the
    blameless path (49~instances) and the blameful path (40~confirmed
    instances after LLM blame attribution and manual validation).}}%
  {%
  
  \begin{tikzpicture}[node distance=3mm and 7mm, every node/.append style={minimum height=8mm}]
    \node[pbox, minimum width=27mm] (bb) {Bug bounty\\findings};
    \node[pbox, below=of bb, minimum width=27mm] (pt) {Pentest\\findings};
    \node[pbox, below=of pt, minimum width=27mm] (is) {Internal security\\findings};
    \node[ptitle, above=2mm of bb] {Sources};
    \node[pbox, right=of pt, minimum width=30mm] (filt) {Apply filtering\\{\footnotesize fix PR present, code change}};
    \node[ptitle] at (filt |- bb.north) [yshift=4mm] {Shared corpus};
    \node[pbox, right=of filt, minimum width=30mm] (bl) {Blameless dataset\\{\footnotesize 49 instances}};
    \node[ptitle] at (bl |- bb.north) [yshift=4mm] {Derived datasets};
    \node[pbox, below=7mm of filt, minimum width=30mm] (ai) {AI blame attribution\\{\footnotesize LLM proposes culprit PR}};
    \node[pbox, right=4mm of ai, minimum width=28mm] (mv) {Manual validation\\{\footnotesize confirmed / rejected}};
    \node[pbox, right=4mm of mv, minimum width=28mm] (bf) {Blameful dataset\\{\footnotesize 40 confirmed instances}};
    \coordinate (join) at ($(filt.west)+(-4mm,0)$);
    \draw[draw=black!70, thick] (bb.east) -| (join);
    \draw[draw=black!70, thick] (is.east) -| (join);
    \draw[parrow] (pt.east) -- (filt.west);
    \draw[parrow] (filt.east) -- (bl.west);
    \draw[parrow] (filt.south) -- (ai.north);
    \draw[parrow] (ai) -- (mv);
    \draw[parrow] (mv) -- (bf);
  \end{tikzpicture}
  
  }
\end{figure}

\subsection{Blameful Dataset Construction}
\label{sec:blameful}

The \textbf{blameful} strategy asks whether the agent can find a vulnerability in the
pull request that introduced it. This is the evaluation setting closest to production because it mirrors exactly what the agent is asked to do in
our CI/CD pipeline: review a pull request before it merges and identify
security issues before they enter the codebase.

Answering this question requires knowing which PR introduced the vulnerability,
information not directly recorded in the ticket. We used an LLM-based skill
built on our internal AI assistant platform that, given the vulnerability
description and the fix~PR diff, proposes the most likely culprit~PR
\citep{sliwerski2005}. The skill reasons about the fix by identifying which
code patterns were removed or reversed, then searches the repository's commit
and PR history for changes that introduced those patterns. It returns a ranked
list of candidate blame~PRs with a confidence rationale for each. We exclude
from the blameful dataset tickets where no plausible culprit~PR could be
identified, whether because the vulnerability predated the PR history, was
introduced through a dependency rather than application code, or because the
fix involved enough refactoring to make reverse attribution ambiguous.

Similarly, some vulnerabilities could not be cleanly attributed to a single
PR. In cases where the vulnerability required a sequence of changes across
multiple pull requests, for example where one~PR introduces vulnerable code
behind a feature flag and a subsequent~PR enables that flag, the blame
attribution is inherently ambiguous since neither~PR alone is sufficient to
introduce the exploitable condition. These cases were labelled as
\emph{partial} during manual validation conducted by one of the authors and
excluded from the blameful dataset. While this exclusion reduces dataset size,
it preserves the integrity of the evaluation: instances where the culprit~PR
is definitively established produce more reliable ground truth than instances
where the introduction is distributed across multiple code changes.

Each instance in the blameful dataset consists of the culprit~PR diff, the
repository context at the merge commit, and the ticket describing the
vulnerability. The vulnerability description is never shown to the security
agent and is reserved exclusively for the judge.

\subsection{Blameless Dataset Construction}
\label{sec:blameless}

The \textbf{blameless} strategy asks whether, given a repository state in which a known
vulnerability already exists, the agent can find it. Rather than presenting
the agent with a diff, we give it access to the repository at the commit
immediately preceding the fix~PR merge, the last state in which the
vulnerability was present in the codebase.

This approach is more scalable than the blameful strategy because it does not
require blame attribution at all. Every ticket in our filtered corpus with a
clean fix~PR reference is a direct candidate, and no LLM-assisted culprit
identification or manual annotation is required. The pre-fix commit hash is
derived from the fix~PR's merge commit parent, and the repository at that hash
becomes the evaluation context. For each candidate instance, an LLM judge
independently checks whether the vulnerability described in the ticket is
detectable in the pre-fix repository state and whether it is no longer
detectable in the post-fix state. Instances that produce a clean yes-to-no
transition are retained as high-confidence blameless instances, while those
that do not yield a clear determination in either state are excluded as
ambiguous. As with the blameful dataset, the ticket is never shown to the
security agent during evaluation and is reserved for use by the judge,
described in detail in Section~\ref{sec:pipeline}.

\subsection{Comparison of the Two Strategies}
\label{sec:comparison}

The blameful strategy is more faithful to the CI/CD setting because the
agent's task of reviewing a PR diff to catch issues before they merge directly
mirrors one of its two primary production roles. The cost of this fidelity is
the blame attribution step, which limits how quickly the dataset can grow.

The blameless strategy reflects the second equally important production role:
scanning existing code for vulnerabilities that were introduced before
automated review was in place, or that somehow bypassed CI/CD gates. Because
quality filtering in the blameless strategy is automated through the temporal
oracle, the dataset can grow more rapidly and can incorporate findings from
sources where blame attribution is not feasible.

Together, the two strategies provide evaluation signals across both dimensions
of what a security agent is expected to do in practice. We recommend that
organisations building similar evaluation infrastructure maintain both
strategies, using blameless evaluation for rapid iteration and prompt
improvement cycles, and reserving blameful evaluation for measuring
performance on the CI/CD interception task specifically.

\section{Evaluation Pipeline}
\label{sec:pipeline}

With a ground-truth dataset in hand, we needed a pipeline that could run our
code review agent against each evaluation instance and automatically score the
result. We built a two-agent pipeline consisting of a
\textbf{Security Engineer} and a \textbf{Judge}. The Security Engineer reviews code without any knowledge of the
underlying vulnerability and produces structured findings. The Judge then
takes those findings alongside the original vulnerability description and
determines whether the Security Engineer's findings, if addressed, would have
substantially mitigated the issue. Figure~\ref{fig:pipeline-eval} illustrates the full pipeline architecture,
including the prompt optimisation loop described in Section~\ref{sec:opro}.
We follow the same architecture for both evaluation strategies, with the
Security Engineer receiving a PR diff in the blameful setting and a pre-fix
repository snapshot in the blameless setting.

\begin{figure}[htbp]
\floatconts
  {fig:pipeline-eval}%
  {\caption{Evaluation pipeline. The Security Engineer agent reviews the code
    under review without knowledge of the underlying vulnerability and produces
    structured SARIF findings. In the blameful strategy the input is a PR diff;
    in the blameless strategy it is a pre-fix repository snapshot. The Judge
    agent evaluates those findings against the vulnerability description and
    returns a yes/no/unknown verdict. An Optimiser LLM examines message traces
    from missed vulnerabilities and proposes targeted improvements to the
    Security Engineer's system prompt.}}%
  {%
  
  \resizebox{\linewidth}{!}{%
  \begin{tikzpicture}[node distance=3.5mm and 6mm]
    \node[pbox, minimum width=62mm] (se) {\textbf{Agent 1: Security Engineer (ReAct)}\\[2pt]
      {\footnotesize Input: code under review, no vulnerability context}\\
      {\footnotesize (PR diff in blameful; repository snapshot in blameless)}\\
      {\footnotesize Tools: read file, list directory, search filenames}\\
      {\footnotesize Output: SARIF JSON findings}};
    \node[pbox, below=of se, fill=white, font=\small\ttfamily] (tr) {security\_agent\_traces.jsonl};
    \node[pbox, below=of tr, minimum width=36mm, xshift=-14mm] (judge) {\textbf{Agent 2: Judge}\\{\footnotesize Output: yes / no / unknown}};
    \node[pbox, right=of judge, minimum width=22mm] (tix) {\textbf{Vulnerability tickets}\\{\footnotesize descriptions}};
    \node[draw=black!50, rounded corners=3pt, fit=(se)(tr)(judge)(tix), inner sep=3mm, inner ysep=5.5mm] (eval) {};
    \node[ptitle, anchor=north west] at ($(eval.north west)+(2mm,-1mm)$) {Evaluation (per instance)};
    \node[pbox, above=6mm of eval, font=\small\ttfamily] (prompt) {llm\_review\_system\_prompt.txt};
    \node[pbox, below=6mm of eval, minimum width=36mm] (score) {\textbf{Score: discovery rate}\\{\footnotesize fraction of instances judged yes}};
    \node[pbox, below=of score, minimum width=62mm] (opt) {\textbf{Optimiser LLM}\\[2pt]
      {\footnotesize Reads traces of missed vulnerabilities}\\
      {\footnotesize Proposes an improved system prompt}};
    \draw[parrow] (prompt) -- node[plabel, right=1.2mm] {prompt $v_n$} (eval.north);
    \draw[parrow] (se) -- node[plabel, right=1.2mm] {SARIF + message trace} (tr);
    \draw[parrow] (tr.south) -- ++(0,-1.5mm) -| node[plabel, pos=0.92, right=1mm] {SARIF} (judge.north);
    \draw[parrow] (tix) -- (judge);
    \draw[parrow] (eval.south) -- node[plabel, right=1.2mm] {verdicts} (score);
    \draw[parrow] (score) -- node[plabel, right=1.2mm] {score + current prompt} (opt);
    \coordinate (R) at ($(eval.east)+(6mm,0)$);
    \coordinate (L) at ($(eval.west)+(-6mm,0)$);
    \draw[parrow] (tr.east) -- (R |- tr.east) |- node[plabel, pos=0.25, right=1.2mm] {message traces} (opt.east);
    \draw[parrow] (opt.west) -- (L |- opt.west) |- node[plabel, pos=0.25, left=1.2mm] {new prompt $v_{n+1}$} (prompt.west);
  \end{tikzpicture}}
  
  }
\end{figure}
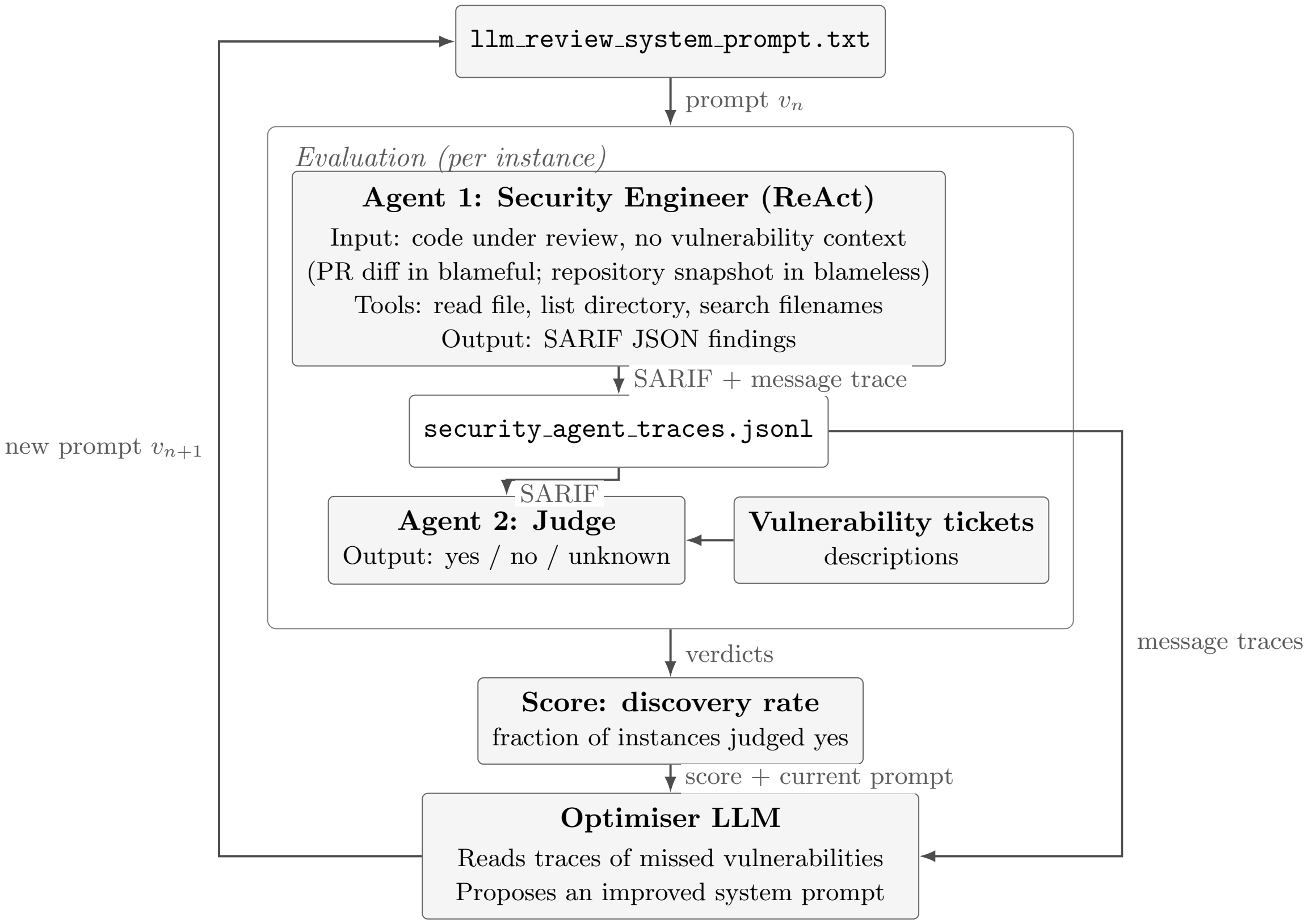

\subsection{The Security Engineer Agent}
\label{sec:se-agent}

The Security Engineer receives the evaluation input, either a culprit PR diff
in the blameful setting or access to a repository snapshot at a pre-fix commit
in the blameless setting, and is asked to identify any security issues it can
find. It operates with no knowledge of the vulnerability it is being evaluated
against (or indeed whether there is a vulnerability at all), never seeing the
ticket or any signal that the code contains a known issue. This blind
condition is what makes the discovery rate meaningful.

The prompt used by the Security Engineer is taken verbatim from our production
LLM-based code review system, the same prompt that runs on real pull requests
in our CI/CD pipeline. This is an intentional design choice so that
improvements measured in the evaluation pipeline translate directly to
improvements in the live system.

The Security Engineer operates in an agentic loop with access to three tools:
reading a file at a specific path in the repository, listing the contents of a
directory, and searching filenames by regex pattern. The loop runs for up to
60~turns per evaluation instance, after which the output is parsed as SARIF
JSON, a structured format for representing security analysis findings. The
complete sequence of tool calls, tool results, and LLM responses for each run
is written to a trace file per evaluation instance.

{\sloppy
We accessed all three models through our internal model gateway. GPT-5.5 was
called through an OpenAI-compatible endpoint, and Sonnet~4.6 and Opus~4.7 were
called through AWS Bedrock using the identifiers
\texttt{global.anthropic.\allowbreak claude-\allowbreak sonnet-\allowbreak 4-6} and
\texttt{global.anthropic.\allowbreak claude-\allowbreak opus-\allowbreak 4-7}. These are provider model
identifiers rather than dated snapshots, so a provider-side update could change
a model's behaviour over time. We did not set temperature, top-p, or a random
seed for any model, which means every run used the provider defaults. Individual
file reads were truncated at 50{,}000 characters, and an instance whose output
failed SARIF validation or raised an error was retried up to two times.\par}

\subsection{The Judge Agent}
\label{sec:judge}

Once the Security Engineer has produced its SARIF output, the Judge receives
two inputs: the SARIF findings and the vulnerability description for that
instance, which the Security Engineer never saw. The Judge's task is to determine
whether the Security Engineer's findings, if acted upon by a developer, would
have substantially mitigated the reported vulnerability. It returns one of
three verdicts: \emph{yes}, \emph{no}, or \emph{unknown}. The Judge was
GPT-5.5 for every Security Engineer model, using the same default sampling
settings.

We use \emph{discovery rate}, the fraction of evaluation instances where the
Judge returns yes, as our primary metric. We track unknown verdicts separately
because a high unknown rate signals a judge reliability problem rather than an
agent performance problem. Discovery rate is a recall measure and it asks whether
the agent surfaced a vulnerability we already know to exist. This framework was
designed to measure recall only and does not measure precision. Our ground truth
is incomplete by construction, since it contains only vulnerabilities that were
reported and confirmed, so a finding that matches no known ticket is not
evidence of a false positive and we do not score it as one.

\subsection{Prompt Optimisation}
\label{sec:opro}

We ran a prompt optimisation experiment using an OPRO-style
approach \citep{yang2024}. A separate LLM, also GPT-5.5, was given the Security Engineer's
traces from instances where the Judge returned no, along with the current
prompt and the security ticket descriptions for those instances, and asked to
propose targeted changes that might address the observed failure patterns.
Because no held-out split was maintained, we limit optimisation to a single
round per evaluator to reduce the risk of in-distribution overfitting.

\subsection{Prompt Design for Each Evaluation Strategy}
\label{sec:prompt-design}

Both prompts share a common output contract: findings are returned as SARIF
JSON with file path, line range, code snippet, and a message containing both a
vulnerability description and remediation steps. Both apply the same severity
classification (HIGH, MEDIUM, LOW) and a confidence scoring floor, where
findings below 0.7 confidence are not reported.

The blameful prompt is designed for PR-time review and scopes the agent
explicitly to security implications newly introduced by the diff under review.
It does not ask the agent to audit pre-existing code unless the PR directly
calls into, extends, enables, or rewires something demonstrably insecure. This
narrowing reflects a practical constraint of the PR review setting, since
surfacing vulnerabilities the PR author did not introduce creates friction and
undermines trust in the tool over time. The blameful prompt applies stricter
false positive filtering, with an explicit list of excluded finding categories
and a strong instruction to prefer missing theoretical issues over generating
noise. The analysis methodology follows three phases: repository context
research, comparative analysis, and vulnerability assessment of the diff.

The blameless prompt is designed for full-codebase audit. The agent is given
access to a fixed set of file paths to audit and can search, browse, and read
the entire repository at a fixed commit. The analysis methodology follows a
different three-phase structure: reconnaissance to map all entry points, sinks,
and trust boundaries; taint analysis tracing data flows from user-controlled
sources to dangerous operations; and an authorisation audit verifying every
entry point has a functioning access control gate. The blameless prompt also
includes a set of worked vulnerability examples with code, covering common
patterns relevant to our organisation.

These structural differences have direct implications for agent behaviour that
we examine in Section~\ref{sec:behavior}. The blameful prompt's narrower scope
and stricter filtering produce more conservative output, while the blameless
prompt's broader mandate and richer examples appear to drive deeper exploration
and more findings per instance.

\FloatBarrier
\section{Experiments and Results}
\label{sec:experiments}

\subsection{Baseline Discovery Rate}
\label{sec:baseline}

We define the baseline as the discovery rate achieved by the Security Engineer
operating with the production prompt verbatim before any improvement.

For the blameful evaluator, the baseline discovery rate with Sonnet~4.6 was
approximately 32\%. For the blameless evaluator, the baseline discovery rate
was approximately 61\% with GPT-5.5.

The blameless evaluator, which was run using GPT-5.5, produced a higher
baseline than the blameful evaluator running Sonnet~4.6, likely reflecting
both a structural difference in task difficulty and a potential model
difference. Disentangling these two factors motivated a more systematic
comparison across models.

\subsection{Model Comparison}
\label{sec:model-comparison}

We expanded our experiments to compare performance across three frontier
models: GPT-5.5, Sonnet~4.6, and Opus~4.7. Running the same production prompt
across all three models on the blameful dataset revealed striking variance.
GPT-5.5 achieved a discovery rate of 52\%, Sonnet~4.6 achieved 32\%, and
Opus~4.7 achieved 18\%. The unknown verdict rates differed as well: Sonnet~4.6
accumulated 7~unknown verdicts, while GPT-5.5 and Opus~4.7 each produced
only~1.

On the blameless dataset, the model ordering shifted. Sonnet~4.6 led at
71\%, followed by GPT-5.5 at 61\% and Opus~4.7 at 37\%. The differences in how models perform across the two strategies is itself a
key finding and is examined in detail in Section~\ref{sec:behavior}.

Table~\ref{tab:ci-blameful} reports 95\%
confidence intervals for all discovery rates, and Table~\ref{tab:significance}
reports the $p$-values for the key comparisons, computed with a chi-square test
of independence on the counts of yes verdicts. Because we report seven such
comparisons, some of them would be expected to fall below 0.05 by chance
alone. Applying a Bonferroni correction gives an adjusted threshold of
$0.05/7 \approx 0.007$, and only the GPT-5.5 versus Opus~4.7 comparison on the
blameful dataset ($p = 0.001$) remains below it. We therefore treat every other
comparison in this paper as exploratory and directional. In the blameful setting, the gap
between GPT-5.5 and Sonnet~4.6 does not reach conventional significance at
this dataset scale ($p = 0.07$) and should be interpreted as a directional
finding; the gap between GPT-5.5 and Opus~4.7, however, is statistically
significant ($p = 0.001$). In the blameless setting, the gap between
Sonnet~4.6 and GPT-5.5 is not statistically significant ($p = 0.29$).
Figure~\ref{fig:baseline-both} shows the verdict distributions for both strategies.

\begin{table}[htbp]
\floatconts
  {tab:ci-blameful}%
  {\caption{95\% confidence intervals for all discovery rates across both strategies.}}%
  {%
\small
\begin{tabular}{lcccc}
\toprule
& \multicolumn{2}{c}{\textbf{Blameful ($n=40$)}}
& \multicolumn{2}{c}{\textbf{Blameless ($n=49$)}} \\
\cmidrule(lr){2-3}\cmidrule(lr){4-5}
\textbf{Model} & \textbf{Rate} & \textbf{95\% CI}
               & \textbf{Rate} & \textbf{95\% CI} \\
\midrule
GPT-5.5 baseline    & 52.5\% & [37.5\%, 67.1\%] & 61.2\% & [47.2\%, 73.6\%] \\
GPT-5.5 post-OPRO   & 57.5\% & [42.2\%, 71.5\%] & 67.3\% & [53.4\%, 78.8\%] \\
Sonnet~4.6 baseline   & 32.5\% & [20.1\%, 48.0\%] & 71.4\% & [57.6\%, 82.2\%] \\
Sonnet~4.6 post-OPRO  & 30.0\% & [18.1\%, 45.4\%] & 69.4\% & [55.5\%, 80.5\%] \\
Opus~4.7 baseline     & 17.5\% & [8.7\%,  32.0\%] & 36.7\% & [24.7\%, 50.7\%] \\
Opus~4.7 post-OPRO    & 17.5\% & [8.7\%,  32.0\%] & 44.9\% & [31.9\%, 58.7\%] \\
\bottomrule
\end{tabular}
  }
\end{table}

\begin{table}[htbp]
\floatconts
  {tab:significance}%
  {\caption{Significance of key comparisons (chi-square test, uncorrected for
  multiple comparisons). With seven comparisons, the Bonferroni-adjusted
  threshold is $0.05/7 \approx 0.007$.}}%
  {%
\small
\begin{tabular}{lc}
\toprule
\textbf{Comparison} & \textbf{$p$-value} \\
\midrule
GPT-5.5 vs.\ Sonnet~4.6 (blameful)    & 0.07  \\
GPT-5.5 vs.\ Opus~4.7 (blameful)      & 0.001 \\
Sonnet~4.6 vs.\ Opus~4.7 (blameful)   & 0.121 \\
GPT-5.5 OPRO improvement (blameful)   & 0.653 \\
Sonnet~4.6 vs.\ GPT-5.5 (blameless)   & 0.285 \\
GPT-5.5 OPRO improvement (blameless)  & 0.527 \\
Opus~4.7 OPRO improvement (blameless) & 0.41  \\
\bottomrule
\end{tabular}
  }
\end{table}

\begin{figure}[htbp]
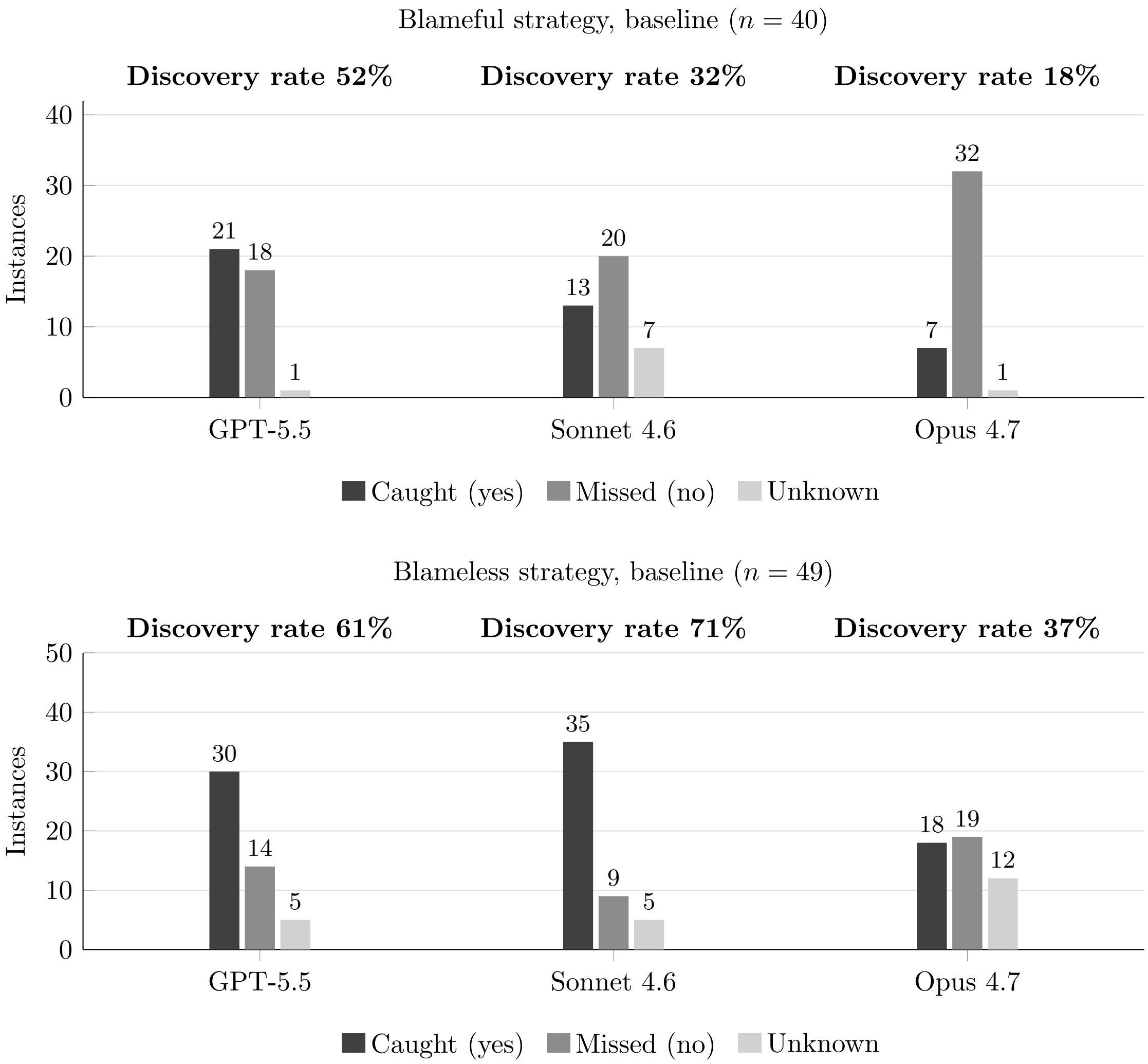

\floatconts
  {fig:baseline-both}%
  {\caption{Baseline verdict distributions before OPRO optimisation.
    \textbf{Top:} Blameful ($n=40$): GPT-5.5 leads at 52\%,
    Sonnet~4.6 at 32\%, Opus~4.7 at 18\%; gap GPT-5.5/Sonnet directional
    ($p=0.07$), GPT-5.5/Opus significant ($p=0.001$).
    \textbf{Bottom:} Blameless ($n=49$): Sonnet~4.6 leads at 71\%,
    GPT-5.5 at 61\%, Opus~4.7 at 37\%; gap Sonnet/GPT-5.5 not significant
    ($p=0.29$).}}%
  {%
  
  \verdictchart{Blameful strategy, baseline ($n=40$)}{42}
   {(GPT-5.5,21) (Sonnet 4.6,13) (Opus 4.7,7)}
   {(GPT-5.5,18) (Sonnet 4.6,20) (Opus 4.7,32)}
   {(GPT-5.5,1) (Sonnet 4.6,7) (Opus 4.7,1)}
   {52\%}{32\%}{18\%}\\[4mm]
  \verdictchart{Blameless strategy, baseline ($n=49$)}{50}
   {(GPT-5.5,30) (Sonnet 4.6,35) (Opus 4.7,18)}
   {(GPT-5.5,14) (Sonnet 4.6,9) (Opus 4.7,19)}
   {(GPT-5.5,5) (Sonnet 4.6,5) (Opus 4.7,12)}
   {61\%}{71\%}{37\%}
  
  }
\end{figure}

\subsection{Prompt Optimisation Results}
\label{sec:opro-results}

On the blameful dataset, GPT-5.5 showed a directional improvement from 52\%
to 57\% after optimisation, though at this dataset scale the difference falls
within run-to-run variance ($p = 0.65$, see Table~\ref{tab:significance}) and
should not be interpreted as a confirmed gain. Sonnet~4.6 and Opus~4.7 did not
meaningfully improve, remaining at approximately 30\% and 18\% respectively.
Figure~\ref{fig:opro-both} shows the post-optimisation distributions for both strategies.

\begin{figure}[htbp]
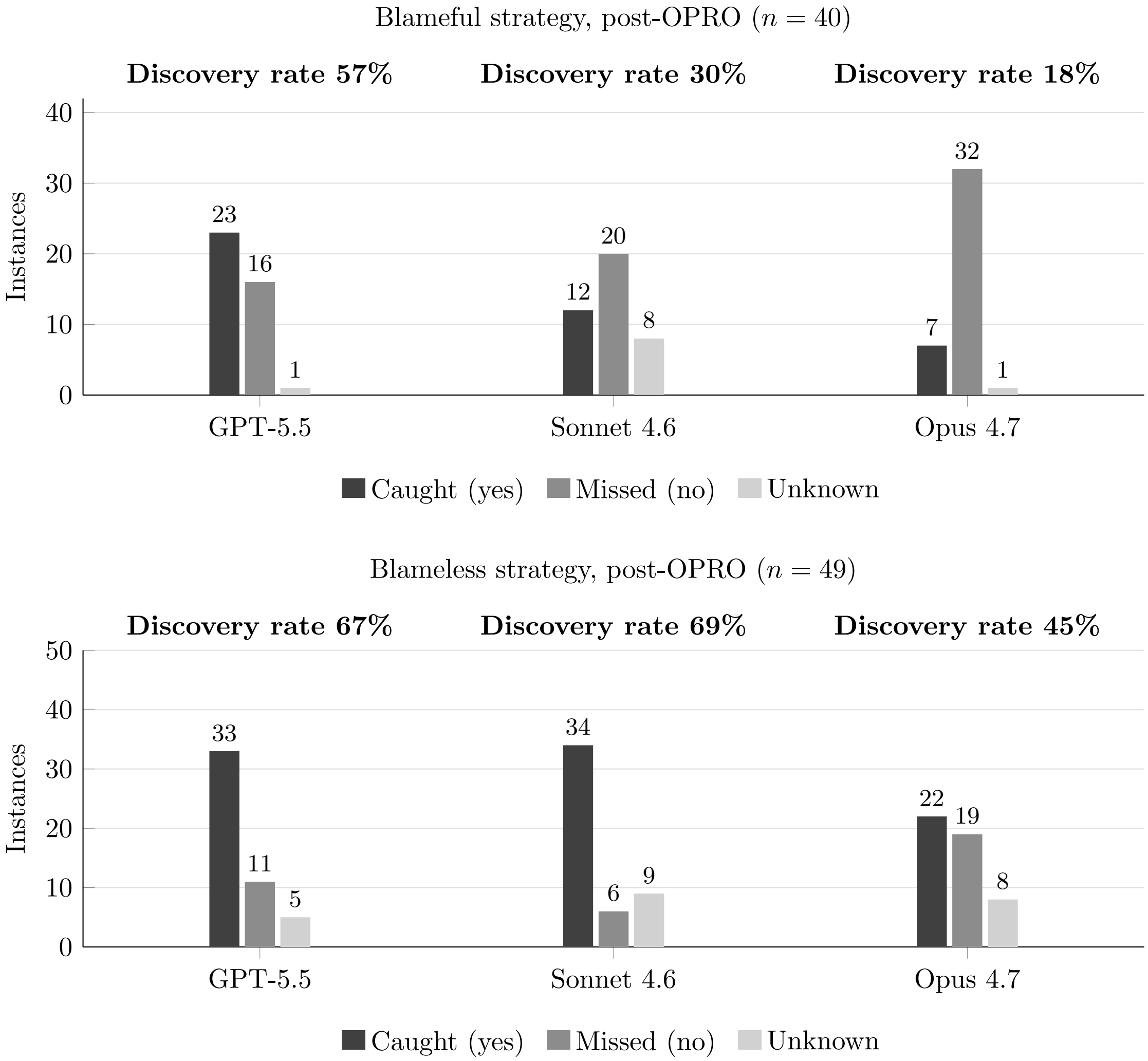

\floatconts
  {fig:opro-both}%
  {\caption{Post-OPRO verdict distributions.
    \textbf{Top:} Blameful ($n=40$): GPT-5.5 improves directionally
    from 52\% to 57\% ($p=0.65$, within variance); Sonnet~4.6 and
    Opus~4.7 remain effectively unchanged.
    \textbf{Bottom:} Blameless ($n=49$): GPT-5.5 and Opus~4.7 show
    directional gains ($p=0.53$ and $p=0.41$); Sonnet~4.6 holds flat
    at 69\%.}}%
  {%
  
  \verdictchart{Blameful strategy, post-OPRO ($n=40$)}{42}
   {(GPT-5.5,23) (Sonnet 4.6,12) (Opus 4.7,7)}
   {(GPT-5.5,16) (Sonnet 4.6,20) (Opus 4.7,32)}
   {(GPT-5.5,1) (Sonnet 4.6,8) (Opus 4.7,1)}
   {57\%}{30\%}{18\%}\\[4mm]
  \verdictchart{Blameless strategy, post-OPRO ($n=49$)}{50}
   {(GPT-5.5,33) (Sonnet 4.6,34) (Opus 4.7,22)}
   {(GPT-5.5,11) (Sonnet 4.6,6) (Opus 4.7,19)}
   {(GPT-5.5,5) (Sonnet 4.6,9) (Opus 4.7,8)}
   {67\%}{69\%}{45\%}
  
  }
\end{figure}

On the blameless dataset, GPT-5.5 directionally improved from 61\% to 67\%
($p = 0.53$), and Opus~4.7 from 37\% to 45\% ($p = 0.41$), but Sonnet~4.6
held effectively flat at 69\% compared to 71\%, a change consistent with
run-to-run variance.

We considered two explanations for why a single round of optimisation did not
produce a confirmed gain, although we did not test either directly. First,
because no held-out split was maintained, we limited optimisation to a single
round to avoid overfitting to the evaluation instances, which gives the
optimiser little room to refine its proposals iteratively. Second, at $n=40$
and $n=49$ a change of five percentage points corresponds to only two or three
instances, which is within the run-to-run variance discussed in
Section~\ref{sec:limitations}, so a modest true improvement would be difficult
to distinguish from noise at this scale.

\subsection{Model Behaviour Across Both Evaluation Strategies}
\label{sec:behavior}

While between-model discovery rate comparisons are subject to the statistical
limitations discussed in Section~\ref{sec:model-comparison}, a separate and
equally important observation emerges from examining how the same model
behaves across the two evaluation strategies. Sonnet~4.6 achieves 32\% on the
blameful task and 71\% on the blameless task. Opus~4.7 moves from 18\% to
45\%. These within-model differences are behavioural observations about how
the same system responds to structurally different tasks, and they raise a
question that outcome scores alone cannot answer: what is it about the task
structure that produces such different behaviour from the same model?

More importantly, this evaluation pipeline exists to inform a production
deployment decision, and the production setting will look different from either
evaluation strategy. Understanding which model is likely to generalise to
production requires understanding why models behave the way they do, not just
what their discovery rate was on a specific dataset \citep{khurana2025}. To
understand the behavioural drivers, we analysed agent traces across all six
model and evaluator combinations. Table~\ref{tab:divergence} summarises the
findings.

\paragraph{Exploration depth across both strategies.}
Our evaluation demonstrated that every model explores approximately twice as
deeply in the blameless setting as in the blameful setting. Sonnet~4.6 goes
from 25.3 average tool calls to 45.6, Opus~4.7 from 5.8 to 16.1, and GPT-5.5
from 11.2 to 23.5. This reflects a structural difference between the two
prompts: the blameful prompt gives the agent a diff to react to, while the
blameless prompt requires a multi-phase audit starting from a list of modified
files, which forces deeper engagement even on instances the agent ultimately
misses. We also measured the
fraction of missed instances where the agent made fewer than 5~tool calls
before stopping, which we call the \emph{early-stop rate}. This rate dropped
from as high as 68\% in blameful mode to 7\% or below across all models in
blameless mode. Figure~\ref{fig:exploration} shows the exploration depth and
early-stop rates across all six model-strategy combinations.

For Opus~4.7, which had the lowest blameful discovery rate of the three, the
exploration pattern is especially revealing. On 21~of its 31~missed blameful
instances it made fewer than 5~tool calls, and on 12~of its 40~total blameful
instances it made zero tool calls entirely, returning an empty findings array
without examining any files. These are not cases where the agent explored and
found nothing, they are cases where the agent decided the diff looked benign
without performing any exploration at all. When the blameless prompt forced
deeper engagement, Opus's discovery rate climbed from 18\% to 45\%,
confirming that the underlying capability exists but is not triggered by the
diff-anchored blameful task. Trace analysis reveals that these zero-tool-call
failures concentrate specifically on small diffs: one-line parameter changes,
new feature flag declarations, build dependency additions, and schema or
config file modifications, precisely the diff types where authorisation
vulnerabilities are most commonly introduced invisibly.

\begin{figure}[htbp]
\floatconts
  {fig:exploration}%
  {\caption{Agent exploration behaviour across both strategies. All three
    models roughly double their tool-call depth in the blameless setting.
    Opus~4.7's 68\% early-stop rate on blameful misses is the primary driver
    of its low blameful discovery rate.}}%
  {%
  
  \begin{tikzpicture}
  \begin{axis}[stratbars, width=0.36\linewidth, title={Average tool calls}, ymax=52,
    legend to name=stratlegend]
  \addplot[fill=blameful]  coordinates {(Sonnet 4.6,25.3) (Opus 4.7,5.8) (GPT-5.5,11.2)};
  \addplot[fill=blameless] coordinates {(Sonnet 4.6,45.6) (Opus 4.7,16.1) (GPT-5.5,23.5)};
  \legend{Blameful, Blameless}
  \end{axis}
  \end{tikzpicture}\hfill
  \begin{tikzpicture}
  \begin{axis}[stratbars, width=0.36\linewidth, title={Average files read}, ymax=36]
  \addplot[fill=blameful]  coordinates {(Sonnet 4.6,23.9) (Opus 4.7,3.8) (GPT-5.5,12.5)};
  \addplot[fill=blameless] coordinates {(Sonnet 4.6,32.0) (Opus 4.7,11.7) (GPT-5.5,21.8)};
  \end{axis}
  \end{tikzpicture}\hfill
  \begin{tikzpicture}
  \begin{axis}[stratbars, width=0.36\linewidth, title={Early-stop rate on misses (\%)}, ymax=76,
    nodes near coords={\pgfmathprintnumber{\pgfplotspointmeta}\%}]
  \addplot[fill=blameful]  coordinates {(Sonnet 4.6,6) (Opus 4.7,68) (GPT-5.5,22)};
  \addplot[fill=blameless] coordinates {(Sonnet 4.6,0) (Opus 4.7,5) (GPT-5.5,7)};
  \end{axis}
  \end{tikzpicture}\\[1mm]
  \pgfplotslegendfromname{stratlegend}
  
  }
\end{figure}
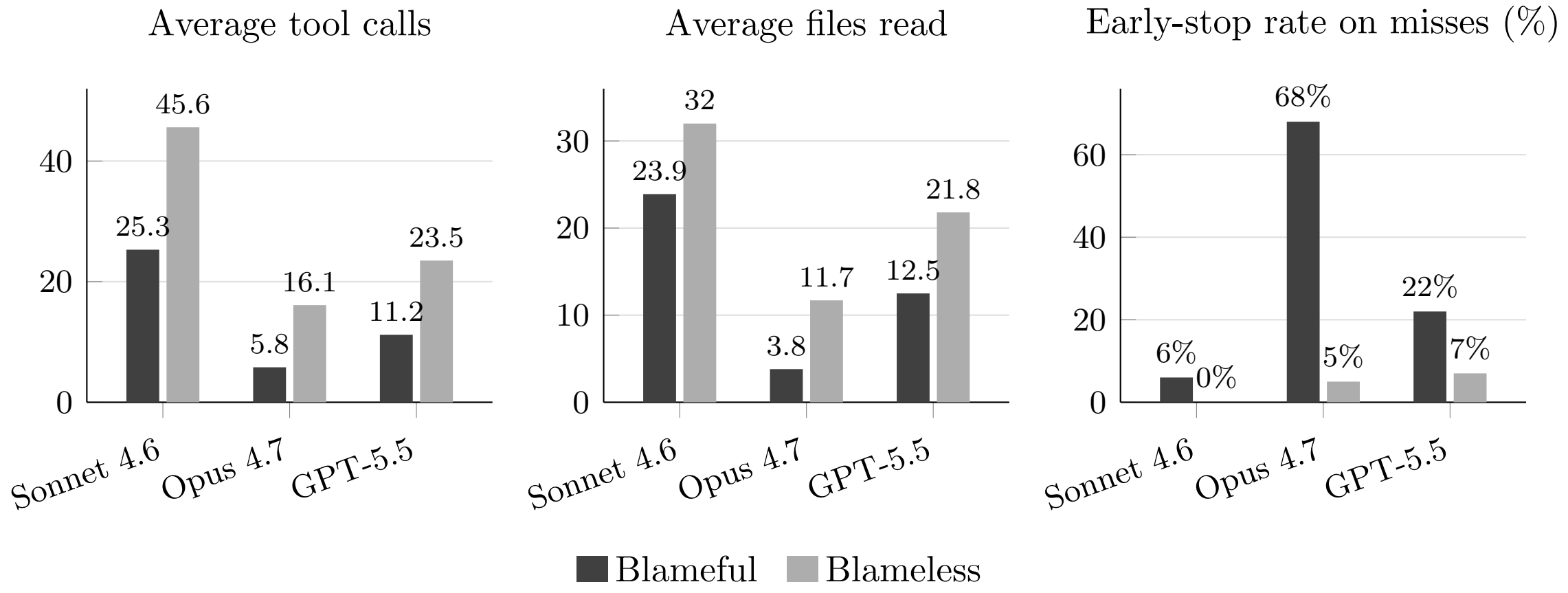

\paragraph{Output quality in the blameful setting.}
Even for models that made a reasonable number of tool calls in the blameful
setting, the most common outcome on a missed instance was producing no security
findings whatsoever: Sonnet~4.6 produced zero findings on 74\% of its blameful
missed instances despite averaging 25.3~tool calls overall. Opus~4.7 produced
zero findings on 90\% and GPT-5.5 on 74\% of their respective blameful missed
instances.

The blameful prompt's structure provides the most plausible account. It
instructs the agent to focus only on vulnerabilities newly introduced by the
diff, applies a strict false positive filtering standard, and explicitly
favours missing theoretical issues over generating noise. A model that finds a
potentially suspicious pattern but cannot fully confirm the taint chain within
the visible files has a strong prompt-level incentive to stay silent.

Yet GPT-5.5 outperforms the other two models in the blameful setting despite
facing the same prompt constraints. Trace analysis of instances where GPT-5.5
caught a vulnerability that Sonnet~4.6 missed reveals two distinct patterns.
The first is \emph{shallow defence acceptance}: Sonnet identifies a
security-relevant pattern, finds something plausibly safe nearby such as a
similar authorisation annotation on a sibling method, and stops its reasoning
there without verifying it actually covers the specific code path being traced.
The second is \emph{lack of attacker mindset}: Sonnet correctly identifies
that code is anomalous but describes its conclusion in code-quality terms
rather than security terms. In one representative case, both models examined
identical code containing a service-level authorisation bypass. Sonnet
characterised it as an unconventional coding approach. GPT-5.5 took the same
observation and reasoned to its security conclusion, that any unauthenticated
caller could trigger a privileged operation on behalf of any user. Both
failure patterns reflect the same underlying gap: shallow security reasoning
depth that falls short of completing the full attack chain.

\paragraph{Output quality in the blameless setting.}
Sonnet~4.6 produced at least some findings on all of its 9~missed blameless
instances, averaging 5.0~findings per missed instance. The agent was finding
real, specific, well-formed security issues with file and line numbers, but
those issues were not the same as the specific one in the security ticket. On
one instance the ticket described a private-message disclosure in a specific
resolution-fetching endpoint, and the agent reported 8~findings covering
missing authorisation checks on five related endpoints in the same service,
all of which could be genuine security issues.

From a production perspective, this is the behaviour you want from a security
agent scanning your production code. A deployed agent should surface all the
real vulnerabilities it finds in a codebase, not just the one a dataset
happens to be scoring it against. The evaluation design scores each instance
against a single vulnerability ticket, but the agent is behaving appropriately for a repository scan, exploring
deeply and reporting real issues with confidence. This is
precisely why trace-level analysis matters alongside outcome scores: an
outcome-only view would lead you to conclude Sonnet~4.6 missed 29\% of
vulnerabilities, while a trace-level view reveals it found multiple real
vulnerabilities on most of those instances, just not the specific one in scope
for that evaluation run.

GPT-5.5, by contrast, exhibits two distinct problems in the blameless setting.
The first is \emph{lateral drift}: rather than deepening analysis of the
primary vulnerable file, it reads adjacent files containing no
security-relevant code and exhausts its tool call budget. The second is a
\emph{systematic blind spot} for certain well-known dangerous API patterns.
Across three separate instances, GPT-5.5 read the exact file containing a
vulnerable YAML deserialisation call and produced no output. The vulnerability
is the use of the SnakeYAML library's default constructor, which does not
restrict the Java types that can be instantiated during deserialisation,
enabling remote code execution via crafted YAML input (CVE-2022-1471, CVSS
9.8~Critical). The preferred alternative is to use a constructor that restricts
type instantiation. Sonnet~4.6, by contrast, identified this pattern
reliably through targeted filename pattern searches, then correctly flagged the
unsafe constructor call upon reading those files. GPT-5.5 read the same files
and produced nothing each time, suggesting a gap in its internalised security
rule set for this specific vulnerability class rather than an exploration
failure.

The differences in model rankings between the two strategies are not primarily
explained by GPT-5.5 performing slightly worse in the blameless setting. It is primarily explained by Sonnet's stronger performance. The blameless prompt is
structurally an exploration-forward task since it explicitly mandates
reconnaissance across the full codebase, targeted searches for known
vulnerability patterns, and multi-phase taint tracing. Sonnet appears to be a
stronger explorer than it is a security reasoner: it excels at broad pattern search across a codebase but struggles
with reasoning about whether a specific code change introduces an exploitable risk. That fundamental difference between the two models is what drives the
difference in performance between the two strategies on our corpus.
Figure~\ref{fig:output-quality} summarises output quality on missed instances
across both settings.

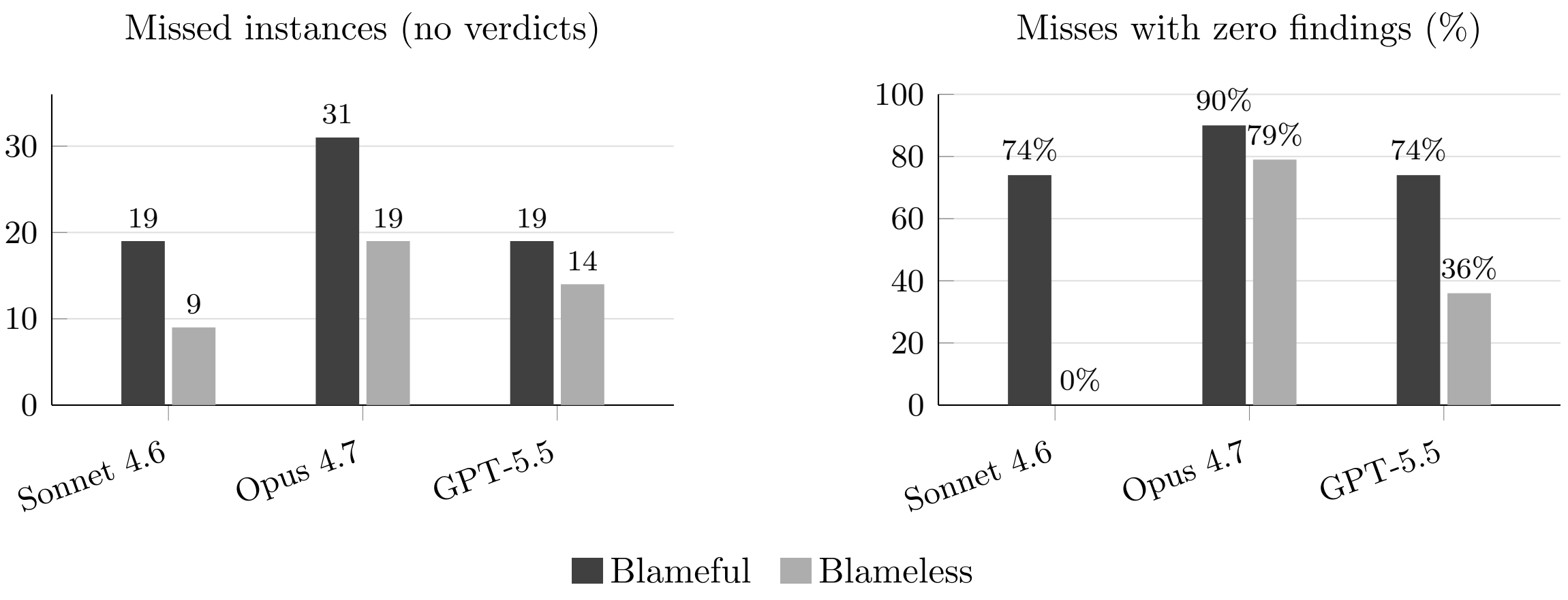
\begin{figure}[htbp]
\floatconts
  {fig:output-quality}%
  {\caption{Output quality on missed instances. Sonnet~4.6 produces zero
    findings on 0\% of blameless misses (it always finds \emph{something},
    just not the evaluated vulnerability), compared to 74\% in blameful mode.
    Opus~4.7's high zero-findings rate persists in both settings.}}%
  {%
  
  \begin{tikzpicture}
  \begin{axis}[stratbars, bar width=12pt, width=0.5\linewidth, title={Missed instances (no verdicts)}, ymax=36,
    legend to name=stratlegendb]
  \addplot[fill=blameful]  coordinates {(Sonnet 4.6,19) (Opus 4.7,31) (GPT-5.5,19)};
  \addplot[fill=blameless] coordinates {(Sonnet 4.6,9) (Opus 4.7,19) (GPT-5.5,14)};
  \legend{Blameful, Blameless}
  \end{axis}
  \end{tikzpicture}\hfill
  \begin{tikzpicture}
  \begin{axis}[stratbars, bar width=12pt, width=0.5\linewidth, title={Misses with zero findings (\%)}, ymax=100,
    nodes near coords={\pgfmathprintnumber{\pgfplotspointmeta}\%}]
  \addplot[fill=blameful]  coordinates {(Sonnet 4.6,74) (Opus 4.7,90) (GPT-5.5,74)};
  \addplot[fill=blameless] coordinates {(Sonnet 4.6,0) (Opus 4.7,79) (GPT-5.5,36)};
  \end{axis}
  \end{tikzpicture}\\[1mm]
  \pgfplotslegendfromname{stratlegendb}
  
  }
\end{figure}

\paragraph{Cross-model blind spots.}
Trace analysis identified four vulnerability classes that all three models
missed consistently. \emph{Business logic and resource exhaustion bugs} were
systematically missed, likely because both the models and the prompt are
heavily optimised toward authorisation-class detection. \emph{Dependency CVEs}
require mapping library version strings to external vulnerability databases, a
capability none of the agents demonstrated without dedicated tooling.
\emph{Field-level data sensitivity bugs} require knowing which response fields
carry internal-only classifications, information not inferable from code
structure alone. \emph{Long upstream taint chains} require tracing
user-controlled inputs through multiple service layers to a dangerous sink,
exceeding the effective reach of all three models. Addressing these blind spots
requires specialised tooling access, fine-tuning on organisation-specific data
classifications, or retrieval-augmented approaches that give the agent access
to CVE feeds, dependency graphs, and internal data sensitivity annotations.

\begin{table}[htbp]
\floatconts
  {tab:divergence}%
  {\caption{Within-strategy divergence summary.}}%
  {%
\small
\begin{tabular}{lll}
\toprule
\textbf{Pattern} & \textbf{Blameful} & \textbf{Blameless} \\
\midrule
Primary miss reason &
  Zero output despite exploration &
  Wrong vulnerability or lateral drift \\
GPT-5.5 advantage &
  Attacker-capability framing &
  N/A \\
Sonnet advantage &
  N/A &
  Deeper targeted search \\
GPT-5.5 gap &
  Conservative filtering &
  Dangerous API blind spot \\
Sonnet gap &
  Code-quality framing &
  Adjacent findings not in scope \\
\bottomrule
\end{tabular}
  }
\end{table}

\paragraph{Takeaways.}
First, the blameful task's diff-anchored structure limits exploration depth
and structurally excludes vulnerability classes that only become apparent when
tracing authorisation patterns across the full service; within that setting
GPT-5.5's attacker-capability framing gives it a consistent advantage.
Second, the blameful prompt's narrower scope and stricter filtering likely
cause models to withhold output on borderline findings, producing conservative
behaviour appropriate for production PR review but masking real capability in
evaluation. Third, on this corpus and with these prompts, GPT-5.5 demonstrated
stronger attacker-capability reasoning while Sonnet~4.6 and Opus~4.7
demonstrated stronger exploratory search behaviour; neither strength dominates
universally. Fourth, blameless evaluation reveals that all models are capable
of producing rich and accurate security findings when given full repository
access, even on instances the evaluation scores as misses, which means
outcome-only scoring can undervalue what these agents can do in
practice and underscores the importance of trace-level behavioural analysis
when selecting a model for production deployment.

\subsection{Limitations}
\label{sec:limitations}

The most significant limitation is dataset scale and composition. With
40~manually annotated culprit~PRs in the blameful dataset and 49~instances in
the blameless dataset, the corpus is small enough that moving a handful of
instances changes the reported discovery rate by several percentage points. Beyond scale, the dataset inherits the discovery
biases of its three source channels: bug bounty programmes, pentest
engagements, and internal engineer findings each tend to surface certain
vulnerability classes over others.

All empirical results come from a single organisation, and we have not
evaluated how well they generalise. The vulnerability classes, languages,
frameworks, and coding conventions in our corpus are those of one organisation,
so the model ordering we report may differ elsewhere, and we present the model
comparisons as observations on this corpus rather than as general rankings. The
methodology also assumes a vulnerability management workflow in which each
confirmed vulnerability is linked to a ticket, a fix pull request, and access
to the repository history. Organisations whose workflows lack any of these
would need to adapt the construction steps, and we have not tested such
adaptations.

Our framework measures recall only. It reports the fraction of known
vulnerabilities that an agent surfaces and was never designed to measure
precision, so it says nothing about the volume or quality of additional findings
the agent produces. Assessing precision would require a different ground truth,
for example labelled findings on code with no known vulnerabilities, and we
leave that to complementary evaluations.

Non-determinism in LLM outputs adds further measurement uncertainty. The same
prompt and model can produce different discovery rates across separate runs on
the same dataset, and the results reported here reflect single runs rather than
averages across multiple executions.

The Judge is GPT-5.5, which is also one of the models under evaluation, so its
verdicts on its own findings may be favourably biased, and we have not measured
this effect.

Finally, judge accuracy is a ceiling on measurement quality. We have not
independently validated the judge against human labels at scale, and the
7~unknown verdicts accumulated by Sonnet~4.6 suggest the judge struggles with
certain finding styles. Building a meta-evaluation of judge accuracy against
human ground truth, and understanding whether the judge applies consistent
scoring standards across different models' finding styles, remain open
problems.

\FloatBarrier
\section{Future Work}
\label{sec:future}

\paragraph{Expanding the evaluation dataset.}
Two complementary approaches are worth exploring. The first is constructing
deliberately insecure pull requests against real production repositories, where
security engineers (human or AI-assisted) introduce known vulnerabilities
without merging them. The culprit~PR is known by construction, eliminating the
blame attribution step, and the fix~PR can be derived by reverting the
vulnerability-introducing commit. The second is constructing a full synthetic
insecure repository purpose-built for evaluation \citep{peng2025}. A larger
dataset would also unlock more rigorous prompt optimisation: once the dataset
grows to a scale where a held-out test split is practical, running multiple
rounds of OPRO with proper train and test separation would allow stronger
generalisation claims. We are already extending the corpus with additional
confirmed vulnerabilities from the same source channels, and we plan to report
results on the larger corpus, together with repeated runs, in follow-up work.

\paragraph{Judge calibration and bias.}
Since models differ substantially in whether they frame findings as attacker
capabilities versus code quality observations, it is worth investigating
whether the judge applies consistent scoring standards across both framing
styles \citep{soumik2026judging}. If the judge penalises findings that
correctly identify a vulnerability but describe it in code-quality terms, then
measured discovery rates are partly a function of finding style rather than
security capability alone. Addressing this requires either a meta-evaluation
of judge verdicts against human security engineer labels, or experiments with
judge prompts that explicitly instruct it to assess the security substance of a
finding independent of how it is framed.

\paragraph{Reducing measurement variance.}
A straightforward improvement is to run each evaluation configuration multiple
times and report discovery rates as means with confidence intervals. A
complementary direction is to investigate whether providing the Security
Engineer prompt with access to more structured tools would reduce variance by
making the agent's search more deterministic.

\paragraph{Generalising the framework.}
Future work should include a structured replication study applying the same
pipeline to a different organisation's operational security data, both to
validate that the methodology generalises and to produce a more diverse
evaluation corpus.

\section{Conclusion}
\label{sec:conclusion}

Building trustworthy evaluation infrastructure for LLM-based security review
benefits from grounding ground truth in real production vulnerabilities, close to
the actual issues your agents will encounter in production. The methodology we
describe is one way to do this, and the blameful and blameless strategies derived
from it give security teams complementary signals across both the PR-review
task and the repository-scanning task. Our experiments show that frontier
models differ meaningfully in the capabilities these strategies measure: some
are stronger at completing the reasoning chain from an anomaly to a concrete
attacker capability, while others are stronger at broad systematic exploration.
Neither profile dominates across both tasks, and the right deployment decision
depends on which task matters most in your environment, what your vulnerability
distribution looks like, and whether chaining models with different strengths
outperforms either alone. We hope this framework gives security teams a
principled foundation for navigating those tradeoffs and choosing the
evaluation strategy, model configuration, and ground truth construction
approach that fits their own codebase and operational reality.

\acks{We thank Mark Adams, who took part in the discussions that shaped this
evaluation approach and who is building tooling on top of the recall
calculation described here, and Jake Brunetto, for serving as a sounding board
throughout this work.}

\bibliography{camlis2026}

\end{document}